\documentclass[aps,prb,twocolumn,showpacs]{revtex4}
\usepackage{graphicx}

\begin{document}


\title{Magnetic-field and current-density distributions in thin-film
superconducting rings and disks}

\author{Ali A. Babaei Brojeny}
\affiliation{%
 Department of Physics, Isfahan University of Technology,
Isfahan 84154, Iran, and Department of Physics and Astronomy,
  Iowa State University, Ames, Iowa, 50011--3160 }
\author{John R.\ Clem}
\affiliation{%
  Ames Laboratory and Department of Physics and Astronomy,\\
  Iowa State University, Ames, Iowa, 50011--3160 }

\date{\today}

\begin{abstract}
We show how to calculate the magnetic-field and sheet-current distributions
for a  thin-film superconducting annular ring (inner radius $a$, outer radius
$b$, and thickness $d<<a$) when either the penetration depth obeys $\lambda <
d/2$ or, if $\lambda > d/2$, the two-dimensional screening length obeys $\Lambda
= 2 \lambda^2/d << a$ for the following cases: 
(a) magnetic flux
$\Phi_z(a)$ trapped in the hole in the absence of an applied magnetic field, 
(b) zero magnetic flux in the hole when the ring is subjected to an applied
magnetic field $H_a$, and 
(c) focusing of magnetic flux into the hole when a magnetic field $H_a$ is
applied but no net current flows around the ring.
We use a similar method to calculate the magnetic-field and sheet-current 
distributions and magnetization loops for a  thin, bulk-pinning-free
superconducting disk (radius $b$) containing a dome of magnetic flux of radius
$a$ when flux entry is impeded by a geometrical barrier.
\end{abstract}

\maketitle

\section{Introduction} 
Recently Babaei Brojeny et al.~\cite{Babaei02} reported exact analytical
solutions for the magnetic-field and sheet-current-density profiles for two
current-carrying parallel coplanar thin-film superconducting strips in a
perpendicular magnetic field.  Included were calculations for (a) the inductance
per unit length when the two strips carry equal and opposite currents, (b) the
zero-flux-quantum state when no net magnetic flux threads between the strips
in a perpendicular applied field $H_a$, and (c) the focusing of
magnetic flux between the two strips in a field $H_a$ when each strip carries no
net current.  These problems are of relevance to the design of superconducting
thin-film devices, especially superconducting quantum interference devices
(SQUIDs).  

Of interest is the focusing of magnetic flux into the central hole in
washer-type~\cite{Ketchen82} SQUIDs and, in particular, the question of how much 
flux $\Phi_h$ goes into the hole when the SQUID is in a perpendicular magnetic
field
$H_a = B_a/\mu_0$ and no net current circulates around the hole.  The
flux-focusing problem was examined by Ketchen
et al.,~\cite{Ketchen85} who expressed $\Phi_h$ in terms of an effective pickup
area of the hole, $A_{eff} = \Phi_h/B_a$, which in general is larger than the
actual area of the hole, $A_h$, but less than the area occupied by the washer,
$A_w$.  
Accounting only for azimuthal currents, they
considered a washer of circular geometry (an annular ring) and derived a simple
theoretical expression for the effective area, $A_{eff} \approx
(8/\pi^2)A_h(A_w/A_h)^{1/2}$, the theoretical approximations used being valid
only for
$A_h << A_w$.  Experiments on a series of square washers with $A_w/A_h$ up to
$~10^4$ yielded results in excellent qualitative agreement with the prediction,
but with 
$A_{eff} \approx
1.1 A_h(A_w/A_h)^{1/2}$.

Experiments by Dantsker et al.~\cite{Dantsker97} on SQUIDs made with narrow
superconducting lines separated by slots or holes (for trapping flux quanta
during cooldown in the earth's magnetic field) have revealed that the presence
of slots or holes increases the effective area over the value for a solid
washer.  This effect was confirmed experimentally by Jansman
et al.,~\cite {Jansman98} who were able to account for the
increased effective area by treating the 
slotted washers as parallel circuits of pickup inductances.

In this paper we
introduce an approach suitable for extension to calculations of the
magnetic-field and sheet-current-density distributions in superconducting
thin-film strips, rings, and narrow lines. 
We consider the idealized case for which the penetration depth $\lambda$ obeys
$\lambda < d/2$ or, if $\lambda > d/2$, the two-dimensional screening length
$\Lambda = 2 \lambda^2/d$ obeys
$\Lambda << a$, such that the key boundary condition is that the normal component
of the magnetic induction is zero on the surface of the superconductor.
A complicating consequence is that the sheet-current distribution in the
superconductor has inverse square-root singularities at the edges.  While a
mutual-inductance approach such as that used by Gilchrist and
Brandt~\cite{Gilchrist96} and Jansman et al.~\cite {Jansman98} is always
applicable, we show here that an approach taking into account the
inverse-square-root singularities from the beginning is simpler and more
efficient.

The authors of Ref.\ ~\onlinecite{Ketchen85} obtained the flux-focusing
result by superposition. 
They first calculated the induced current flowing in the clockwise direction in
an applied magnetic induction $B_a$ assuming zero magnetic flux in the hole. 
They approximated this current using the known result for a superconducting disk
with no central hole.
They next calculated the induced current flowing in the counterclockwise
direction in the absence of an applied field assuming a given amount of magnetic
flux $\Phi_h$ in the hole.  They approximated this current using the known result
for an infinite superconducting sheet with a round hole in it.
Finally they obtained the relation between $B_a$ and $\Phi_h$ by equating the
magnitudes of the two circulating currents.
In the present paper, we show how to calculate all properties without 
making the small-hole approximations
used in Ref.\ ~\onlinecite{Ketchen85}.
We show how to solve the flux-focusing problem directly, as well as by
superposition. 

Another problem of interest is the calculation of the magnetic-field and
current-density distribution for the case of a bulk-pinning-free
type-II superconducting disk of radius $b$ and thickness $d << b$ in which
the entry of magnetic flux is impeded by a geometrical
barrier.~\cite{Schuster94,Zeldov94a}  Analytic solutions for the field and
current distributions and the magnetization in strips subject
to a geometrical barrier have been studied for the  bulk-pinning-free case in
Refs.\ ~\onlinecite{Benkraouda96} and ~\onlinecite{Mawatari03} and for the case
of Bean-model bulk pinning ($J_c$ = const) in Ref.\ ~\onlinecite{Maksimov98}. 
Numerical results for the field and current distributions and the magnetization
in disks subject to both a geometrical barrier and bulk pinning with a
$B$-dependent $J_c$ have been presented in Ref.\ ~\onlinecite{Doyle97}.
In the following, we present an efficient method for calculating the field and
current distributions and the magnetization in bulk-pinning-free disks subject
to a geometrical barrier.  

Our paper is organized as follows.  In Sec.\ II, we outline our approach and set
down the basic equations.  In Sec.\ III, we apply this approach to calculate the
inductance of an annular ring of arbitrary inner radius.   In Sec.\ IV, we
calculate the current circulating around a ring remaining in the
zero-flux-quantum state while subjected to  a perpendicular magnetic field.  
In Sec.\ V, we consider the flux-focusing problem and calculate the
magnetic flux contained in the center of a ring in an applied magnetic
field when there is no net current around the ring.  In Sec.\ VI, we calculate
the magnetization loop for a bulk-pinning-free thin-film type-II superconducting
disk subject to a geometrical barrier.  We briefly discuss our results in Sec.\
VII.

\section{Basic equations} 
We consider a thin-film superconducting annular ring in the plane $z
= 0$, centered on the $z$ axis, with inner and outer radii
$a$ and $b$ and thickness $d << a$.  We assume that either $\lambda < d/2$ or
$\Lambda << a$ if $\lambda > d/2$, as discussed in the introduction.  By the
Biot-Savart law, the $z$ component of the magnetic field in the  plane $z=0$ 
is~\cite{Landau60,Jackson62}
\begin{equation}
  H_z(\rho)=H_a + \frac{1}{2 \pi}\int_a^b G(\rho,\rho')K_\phi(\rho')d\rho',
\label{Hz}
\end{equation}
where $H_a$ is the applied field, $K_\phi(\rho)$ is the sheet-current density
in the counterclockwise direction,  
\begin{equation}
G(\rho,\rho')=K(k)/(\rho+\rho')-E(k)/(\rho-\rho'),
\label{G}
\end{equation}
\begin{equation}
k=2(\rho \rho')^{1/2}/(\rho+\rho'),
\label{k}
\end{equation}
and $K$ and $E$ are complete elliptic integrals of the first 
and second kind with modulus $k$.  
An important boundary condition we will use in this paper is that
$H_z(\rho)=0$ for $a<\rho<b$.
The total current in the counterclockwise
direction is
\begin{equation}
  I=\int_a^b K_\phi(\rho)d\rho,
\label{I}
\end{equation}
and the magnetic moment along the $z$ direction is
\begin{equation}
  m_z=\pi\int_a^b \rho^2 K_\phi(\rho)d\rho.
\label{mz}
\end{equation}
Another quantity of interest is the magnetic flux up through a circle of radius
$\rho$ in the plane $z=0$,~\cite{Landau60,Jackson62}
\begin{equation}
  \Phi_z(\rho)=\mu_0H_a\pi\rho^2 + \frac{\mu_0}{2}\int_a^b
G_A(\rho,\rho')K_\phi(\rho')d\rho',
\label{Phiz}
\end{equation}
where 
\begin{equation}
G_A(\rho,\rho')=(\rho+\rho')[(2-k^2)K(k)-2E(k)]
\label{GA}
\end{equation}
and $k$ is given in Eq. (\ref{k}).

In the following sections we present solutions of the above equations and
determine the corresponding sheet-current density $K_\phi(\rho)$ for four cases:
(a) self-inductance
$L =
\Phi_z(a)/I$ when
$H_a=0$, (b) the zero-flux-quantum state [$\Phi_z(a)=0$] in an applied field
$H_a$, (c) flux focusing in an applied field [calculation of $\Phi_z(a)$ when
$I=0$], and (d) geometrical-barrier effects in a thin disk of radius $b$
containing a Lorentz-force-free magnetic-flux dome of radius $a$.  In each case,
we assume a spatial dependence of the reduced sheet-current density of the form
\begin{equation}
  \tilde{K}_\phi(u)=\frac{4g(u)}{\pi u \sqrt{(u^2-\tilde{a}^2)(1-u^2)}},
\label{K}
\end{equation}
where $u=\rho/b$ and $\tilde{a}=a/b$ and $g(u)$ is a polynomial containing $N$
terms,
\begin{equation}
  g(u) = \sum_{m=1}^{N} g_m (\frac{u-\tilde{a}}{1-\tilde{a}})^{m-1}.
\label{g}
\end{equation}
Although we are not certain that such a choice gives an exact solution in
general, it reduces to known exact solutions in various limits [$a \rightarrow
0, b
\rightarrow
\infty$, or $(b-a) << b$], all of which have inverse-square-root singularities at
the sample edges. 
To determine the $N$ coefficients, $N-1$ equations
 are obtained by setting
$H_z(\rho_n) = 0$, where $\rho_n = a +n(b-a)/N$ and $n = 1, 2, ..., N-1$.  The
$N$-th equation depends on the case under consideration; for case (a) we
use Eq. (\ref{I}) for given $I$, for case (b) we use Eq. (\ref{Phiz})
and set
$\Phi_z(a)=0$, for case (c) we use Eq. (\ref{I}) and  set $I=0$, and for case (d)
we use Eq. (\ref{g}) and set $g(\tilde{a}) = 0$.

For numerical evaluation of the integrals in Eqs.\ (\ref{Hz}), (\ref{I}),
(\ref{mz}), and (\ref{Phiz}), it is convenient to change variables using the
substitution 
$v = \rho'/b = \sqrt{\tilde{a}^2 +(1-\tilde{a}^2)\sin^2\phi}$
and to define the functions 
\begin{equation}
  h_m(u) = \frac{2}{\pi^2} \int_0^{\pi/2} G(u,v)
(\frac{v-\tilde{a}}{1-\tilde{a}})^{m-1}v^{-2} d\phi,
\label{hm}
\end{equation} 
\begin{equation}
  i_m = \frac{4}{\pi}\int_0^{\pi/2}
(\frac{v-\tilde{a}}{1-\tilde{a}})^{m-1}v^{-2} d\phi,
\label{im}
\end{equation}
\begin{equation}
  f_m = \frac{4}{\pi}\int_0^{\pi/2}
(\frac{v-\tilde{a}}{1-\tilde{a}})^{m-1} d\phi,
\label{fm}
\end{equation}
\begin{equation}
  \phi_m(u) = \frac{2}{\pi} \int_0^{\pi/2} G_A(u,v)
(\frac{v-\tilde{a}}{1-\tilde{a}})^{m-1}v^{-2} d\phi,
\label{phim}
\end{equation} 
and 
\begin{equation}
  \alpha_{nm} =  h_m(u_n),
\label{alphanm}
\end{equation}
where $u_n = \rho_n/b = \tilde{a} + n(1-\tilde{a})/N$, and $n = 1, 2, ..., N-1$.
For $a < \rho < b$ ($\tilde{a} < u < 1$), Eqs. (\ref{hm}) and (\ref{phim}) are
principal-value integrals, evaluated by splitting the $\phi$ integral into
two parts, one from 0 to
$\Phi(u-\epsilon)$ and the other from $\Phi(u+\epsilon)$ to $\pi/2$, where
\begin{equation}
  \Phi(u) = \sin^{-1}\sqrt{\frac{u^2-\tilde{a}^2}{1-\tilde{a}^2}}
\label{Phi}
\end{equation}
and $\epsilon$ is an infinitesimal.
For the results presented here we have used $\epsilon = 10^{-7}.$

\section{Inductance of an annular ring} 
To calculate the inductance, we set $H_a=0$ in Eq. (\ref{Hz}) and
define $K_{I\phi} = (I_I/b)\tilde{K}_{I\phi}$, where the subscript $I$
henceforth labels all  quantities that are specific to calculations of the
inductance.  To evaluate the coefficients $g_{Im}$ in 
\begin{equation}
  g_I(u) = \sum_{m=1}^{N} g_{Im}(\frac{u-\tilde{a}}{1-\tilde{a}})^{m-1},
\label{gI}
\end{equation}
we use the $N$ equations
\begin{equation}
  \sum_{m=1}^{N} \alpha_{Inm}g_{Im} =  \beta_{In},
\label{alphaIsum}
\end{equation}
$n = 1, 2, ..., N$,
where $\alpha_{Inm} =  \alpha_{nm}$ and $\beta_{In} = 0$
for  $n<N$, and 
 $\alpha_{INm} =  i_m$ and $\beta_{IN} = 1$ for $n=N.$
These equations  are
obtained from  Eqs.\ (\ref{Hz}), (\ref{K}), (\ref{g}), (\ref{hm}), and
(\ref{alphanm}) and
$H_z(\rho_n) = 0$  for
$n<N$, and  from  Eqs.\ (\ref{I}),  (\ref{K}), (\ref{g}), and (\ref{im}) for
$n=N$.

Numerical results for $\tilde{H}_{Iz} = b H_{Iz}/I_I$, $\tilde{K}_{I\phi}$, and
$g_I$ vs
$u=\rho/b$ for $a = b/2$ $(\tilde{a} = 0.5)$ are shown in Fig.\ 1.
In this calculation, as well as in all others in Secs. III-VI, the magnitude of
the reduced magnetic field for $a < \rho < b$ was less than $10^{-5}$ except for
$\rho$ very close to $a$ or $b$, where the numerical results for the
principal-value integrals in Eq.\ (\ref{hm}) became less accurate. 
\begin{figure}
\includegraphics[width=8cm]{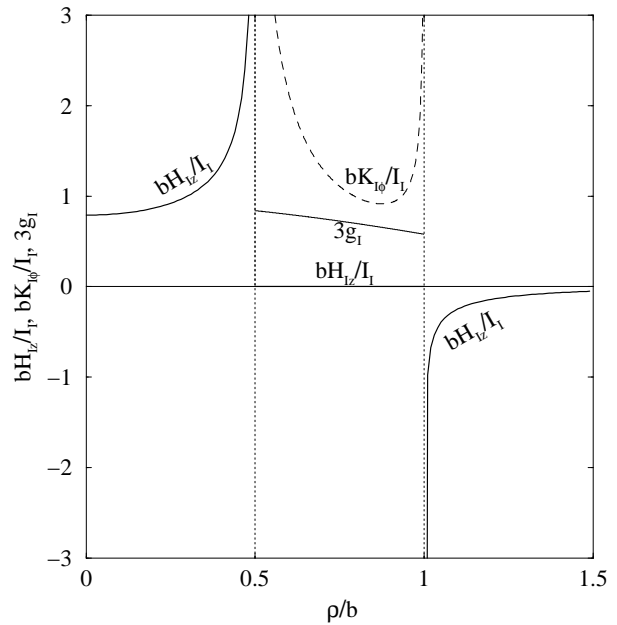}
\caption{%
Reduced magnetic field $\tilde{H}_{Iz} = b H_{Iz}/I_I$,
reduced sheet-current density $\tilde{K}_{I\phi}$, and polynomial
$g_I$ (multiplied by 3) vs
$u=\rho/b$ obtained while calculating the inductance for a superconducting ring
with $\tilde{a} = a/b = 0.5$.}
\label{Fig1}
\end{figure}
Results for
$g_{Im}$ vs
$\tilde{a}$ are shown in Fig.\ 2.  
The inductance is calculated from 
\begin{equation}
  L=\Phi_{Iz}/I_I=\mu_0 b \tilde{\Phi}_{Iz}(\tilde{a}),
\label{L}
\end{equation}
where
\begin{equation}
  \tilde{\Phi}_{Iz}(u)=\sum_{m=1}^{N} \phi_m(u) g_{Im},
\label{PhiIz}
\end{equation}
and is shown in Fig.\ 3 as a function of $\tilde{a} = a/b$.  Dashed lines
in Fig.\ 3 show
expressions valid in the limits of small and large $\tilde{a}$:  For 
$\tilde{a} << 1$, the inductance approaches $L_0 = 2 \mu_0 a$ [or $
\tilde{\Phi}_{Iz}(\tilde{a}) = 2\tilde{a}$], as obtained by Ketchen at
al.,~\cite{Ketchen85}  and for $\tilde{a}
\rightarrow 1$, the inductance approaches
\begin{equation}  
L_1 = \mu_0 R [\ln (8R/w) - (2-\ln4)],
\label{L1}
\end{equation} 
as obtained by Brandt~\cite{Brandt97} for
a superconducting annulus of mean radius $R$ and width $w << R$.  [Here $R =
(a+b)/2 = b(1+\tilde{a})/2$ and $w = (b-a) = b(1-\tilde{a})$.]
Equation (\ref{L1}) can be obtained from~\cite{Poole95} $L = \mu_0 R [\ln (8R/r)
- 2]$, the inductance of a superconducting ring of radius $R$ and wire radius
$r << R$, by replacing $r$ by $w/4$.~\cite{footnote}
The empirical formula
\begin{equation}  
L_2 = \mu_0 b [\tilde{a} -0.197 \tilde{a}^2 - 0.031 \tilde{a}^6
+(1+\tilde{a}) \tanh^{-1}\tilde{a}],
\label{L2}
\end{equation} 
where $\tilde{a} = a/b$, fits our numerical results for $L$ within 0.06\%, and a
plot of it is indistinguishable from the solid curve in Fig.\ 3.

\begin{figure}
\includegraphics[width=8cm]{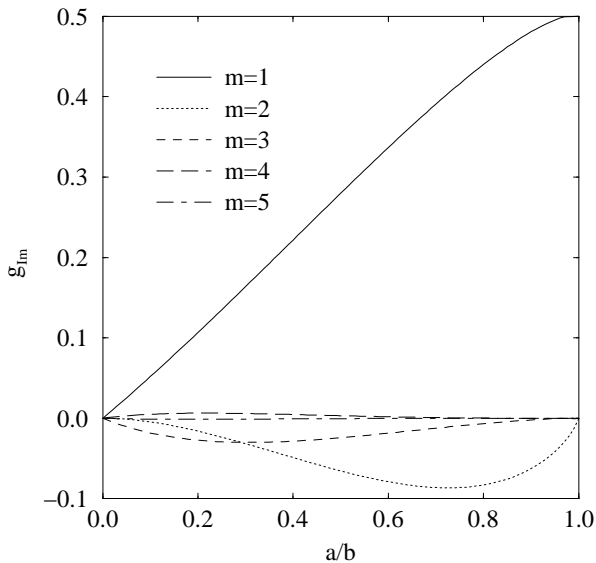}
\caption{%
Coefficients $g_{Im}$ in the polynomial of Eq.\ (\ref{gI}) vs $\tilde{a} =
a/b$ obtained while calculating the inductance of a superconducting ring.}
\label{Fig2}
\end{figure}

The magnetic moment associated with the circulating current can be calculated
from 
\begin{equation}
  m_{Iz}=I_I \pi b^2 \sum_{m=1}^{N} f_m g_{Im}.
\label{mIz}
\end{equation}

\begin{figure}
\includegraphics[angle=270,width=8cm]{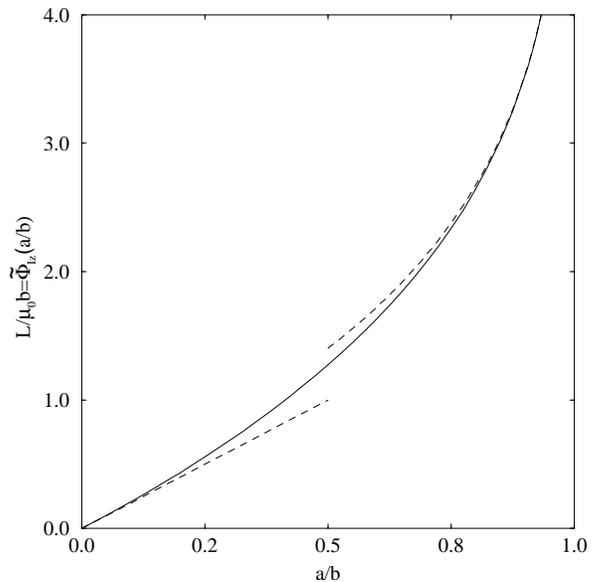}
\caption{%
Reduced inductance of a superconducting ring, $L/\mu_0b$ vs $\tilde{a}
= a/b$, calculated from Eqs.\ (\ref{L}) and (\ref{PhiIz}).  Dashed curves show
approximations valid in the limits
$\tilde{a}
\rightarrow 0$ and $\tilde{a} \rightarrow 1.$}
\label{Fig3}
\end{figure}

\section{Zero-flux-quantum state} 
Consider an annular ring that has been cooled into the superconducting
state in the absence of a magnetic field, such that no magnetic flux
is trapped anywhere in the ring.  When a perpendicular magnetic field
$H_a$ is applied, a circulating current is induced, but the ring remains in the
Meissner state, and the magnetic flux up through the hole remains zero (there
are no flux quanta in the hole). The induced sheet current density is
$K_{Z\phi} = H_a \tilde{K}_{Z\phi}$, where the subscript $Z$ henceforth labels
all quantitites that are specific to calculations for the zero-flux-quantum
state. To evaluate the coefficients $g_{Zm}$ in 
\begin{equation}
  g_Z(u) = \sum_{m=1}^{N} g_{Zm}(\frac{u-\tilde{a}}{1-\tilde{a}})^{m-1},
\label{gZ}
\end{equation}
we use the $N$ equations
\begin{equation}
  \sum_{m=1}^{N} \alpha_{Znm}g_{Zm} =  \beta_{Zn},
\label{alphaZsum}
\end{equation}
$n = 1, 2, ..., N$,
where $\alpha_{Znm} =  \alpha_{nm}$ and $\beta_{Zn} = -1$
for  $n<N$, and 
 $\alpha_{ZNm} =  \phi_m(\tilde{a})$ and $\beta_{ZN} = -\pi \tilde{a}^2$ for
$n=N.$ These equations  are
obtained from  Eqs.\ (\ref{Hz}), (\ref{K}), (\ref{g}), (\ref{hm}), and
(\ref{alphanm}) and
$H_z(\rho_n) = 0$  for
$n<N$, and  from  Eqs.\ (\ref{Phiz}),  (\ref{K}), (\ref{g}), and (\ref{phim}) for
$n=N$.

Numerical results for $\tilde{H}_{Zz} = H_{Zz}/H_a, \tilde{K}_{Z\phi},$ and
$g_Z$ vs $u = \rho/b$ for  $a = b/2$ $(\tilde{a} = 0.5)$ are shown in Fig.\ 4. 
Results for $g_{Zm}$ vs $\tilde{a}$ are shown in Fig.\ 5.  The magnitude  $|I_Z|$
of the induced current is
obtained from
$I_Z = H_a b
\tilde{I}_Z$, where 
\begin{equation}
  \tilde{I}_Z=\sum_{m=1}^{N} i_m g_{Zm},
\label{IZ}
\end{equation}
and is shown in Fig.\ 6 as a function of $\tilde{a} = a/b$.
Dashed lines
in Fig.\ 6 show
expressions valid in the limits of small and large $\tilde{a}$:
For 
$\tilde{a} << 1$, the induced current  approaches $I_Z =  -4 H_a b/\pi$ [or
$\tilde{I}_Z=-4/\pi$], as obtained by Ketchen at al.,~\cite{Ketchen85} and for 
$\tilde{a}
\rightarrow 1$, the induced current  approaches
$I_Z = -\pi R^2 B_a/L_1.$

\begin{figure}
\includegraphics[width=8cm]{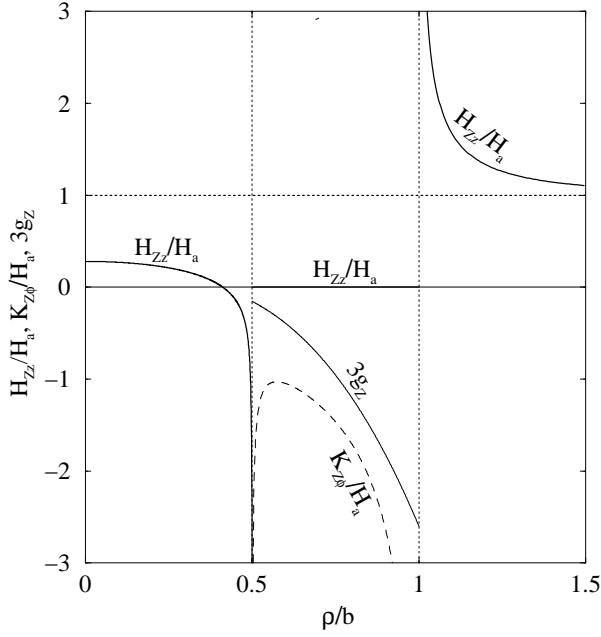}
\caption{%
Reduced magnetic field $\tilde{H}_{Zz} = H_{Zz}/H_a$,
reduced sheet-current density $\tilde{K}_{Z\phi}$, and polynomial
$g_Z$ (multiplied by 3) vs
$u=\rho/b$ for the zero-flux-quantum state
with $\tilde{a} = a/b = 0.5$.}
\label{Fig4}
\end{figure}

\begin{figure}
\includegraphics[width=8cm]{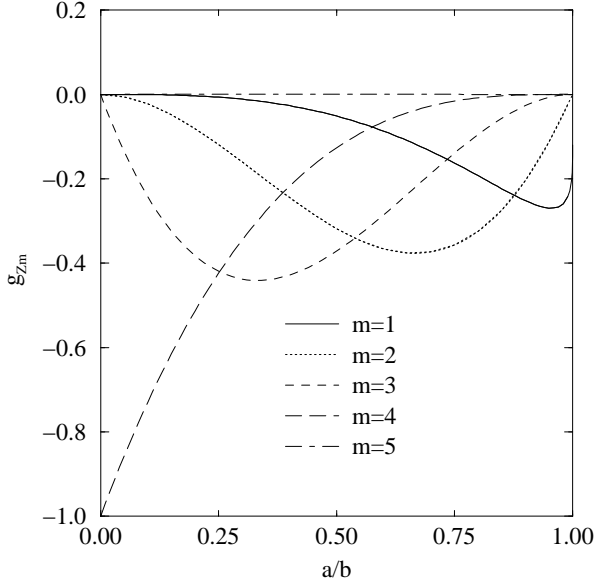}
\caption{%
Coefficients $g_{Zm}$ in the polynomial of Eq.\ (\ref{gZ}) vs $\tilde{a} =
a/b$ for the zero-flux-quantum state.}
\label{Fig5}
\end{figure}

\begin{figure}
\includegraphics[width=8cm]{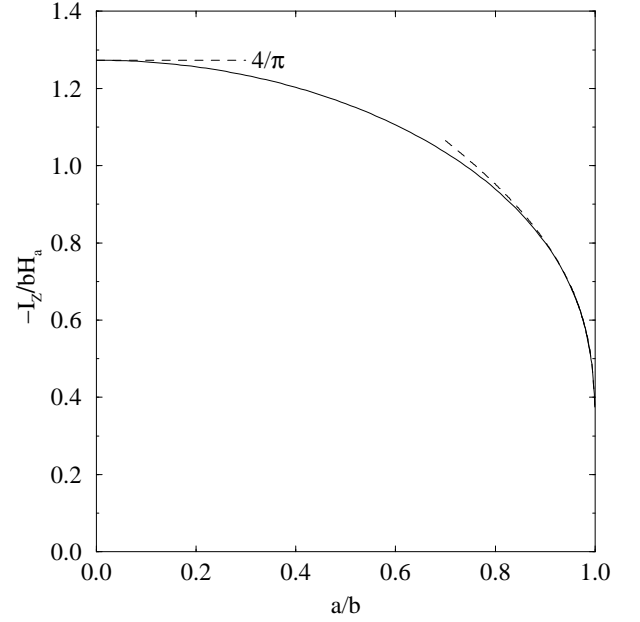}
\caption{%
Magnitude of the reduced current, $-I_Z/bH_a$, vs
$\tilde{a} = a/b$ for the zero-flux-quantum state calculated from Eq.\
(\ref{IZ}).  Dashed curves show
approximations valid in the limits
$\tilde{a}
\rightarrow 0$ and $\tilde{a} \rightarrow 1.$}
\label{Fig6}
\end{figure}

\section{Flux focusing} 

We now solve for the current and field distribution when a superconducting
annular ring is placed in a perpendicular magnetic field $H_a$ subject to the
condition that there is no net current circulating around the ring.  We wish to
determine how much magnetic flux is focused into the hole in the middle of the
ring.  The sheet current density in this case is $K_{F\phi} = H_a
\tilde{K}_{F\phi}$, where the subscript $F$ henceforth labels all quantities
that are specific to calculations of flux focusing.  To evaluate the
coefficients $g_{Fm}$ in
\begin{equation}
  g_F(u) = \sum_{m=1}^{N} g_{Fm}(\frac{u-\tilde{a}}{1-\tilde{a}})^{m-1},
\label{gF}
\end{equation}
we use the $N$ equations
\begin{equation}
  \sum_{m=1}^{N} \alpha_{Fnm}g_{Fm} =  \beta_{Fn},
\label{alphaFsum}
\end{equation}
$n = 1, 2, ..., N$,
where $\alpha_{Fnm} =  \alpha_{nm}$ and $\beta_{Fn} = -1$
for  $n<N$, and 
 $\alpha_{FNm} =  i_m$ and $\beta_{FN} = 0$ for
$n=N.$ These equations  are
obtained from  Eqs.\ (\ref{Hz}), (\ref{K}), (\ref{g}), (\ref{hm}), and
(\ref{alphanm}) and
$H_z(\rho_n) = 0$  for
$n<N$, and  from  Eqs.\ (\ref{I}),  (\ref{K}), (\ref{g}), (\ref{im}), and $I=0$
for
$n=N$.

Numerical results for $\tilde{H}_{Fz} = H_{Fz}/H_a, \tilde{K}_{F\phi},$ and
$g_F$ vs $u = \rho/b$ for  $a = b/2$ $(\tilde{a} = 0.5)$ are shown in Fig.\ 7. 
Results for $g_{Fm}$ vs $\tilde{a}$ are shown in Fig.\ 8.  The magnetic flux
focused into the hole is $\Phi_{Fz}(a) = \mu_o H_a b^2
\tilde{\Phi}_{Fz}(\tilde{a})$, where 
\begin{equation}
\tilde{\Phi}_{Fz}(\tilde{a}) = \pi \tilde{a}^2 +
\sum_{m=1}^{N} \phi_m(\tilde{a}) g_{Fm}.
\label{PhiFz}
\end{equation}

\begin{figure}
\includegraphics[width=8cm]{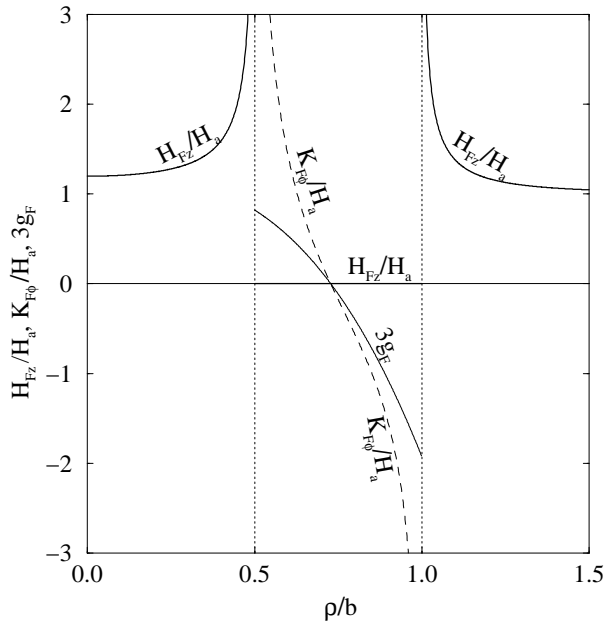}
\caption{%
Reduced magnetic field $\tilde{H}_{Fz} = H_{Fz}/H_a$,
reduced sheet-current density $\tilde{K}_{F\phi}$, and polynomial
$g_F$ (multiplied by 3) vs
$u=\rho/b$ for flux focusing
with $\tilde{a} = a/b = 0.5$.}
\label{Fig7}
\end{figure}

\begin{figure}
\includegraphics[width=8cm]{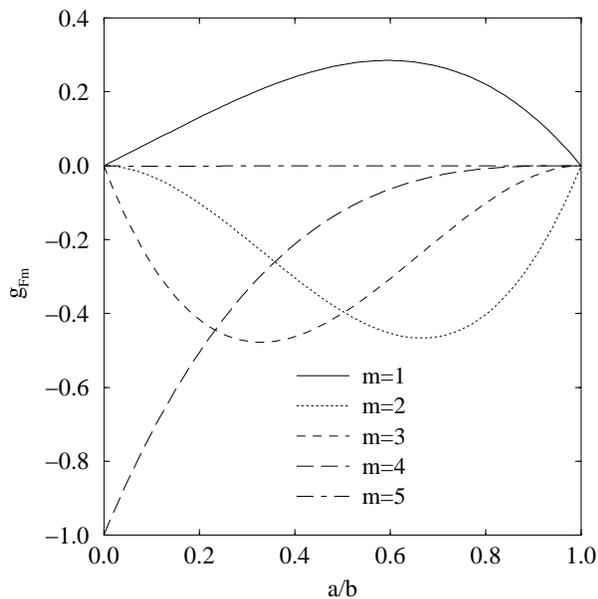}
\caption{%
Coefficients $g_{Fm}$ in the polynomial of Eq.\ (\ref{gF}) vs $\tilde{a} =
a/b$ for flux focusing.}
\label{Fig8}
\end{figure}

The effective area of the hole (which corresponds to the effective pickup area
of a SQUID made of a circular washer), defined via $\Phi_{Fz}(a) = \mu_0 H_a
A_{eff}$, is always larger than the actual area of the hole, $A_h = \pi a^2.$ 
We find
\begin{equation}
\frac{A_{eff}}{A_h} = \frac{\Phi_{Fz}(a)}{\mu_o H_a \pi a^2} = 1 +\frac{1}{\pi
\tilde{a}^2}\sum_{m=1}^{N} \phi_m(\tilde{a}) g_{Fm},
\label{Aeff}
\end{equation}
which is shown in Fig.\ 9 as a function of $1/\tilde{a} = b/a$.  Dashed lines in
Fig.\ 9 show
expressions valid in the limits of small and large $\tilde{a}$:
For 
$\tilde{a} << 1$, $A_{eff}/A_h$  approaches $(8/\pi^2)(b/a)$ [or $8/\pi^2
\tilde{a}$], as obtained by Ketchen at al.,~\cite{Ketchen85} and for 
$\tilde{a}
\rightarrow 1$, $A_{eff}/A_h$   approaches $(R/a)^2$
[or$((1+\tilde{a})/2\tilde{a})^2$], where
$R = (a+b)/2$ is the mean radius of the ring.

The flux-focusing problem also can be solved from a linear superposition of the
fields calculated in Secs. III and IV.  From 
$K_{F\phi}=K_{I\phi}+K_{Z\phi}$ and the condition $I_F = I_I + I_Z = 0$ we obtain
$g_F(u) = -\tilde{I}_Z g_I(u) + g_Z(u)$,  
$g_{Fm} = -\tilde{I}_Z g_{Im} + g_{Zm}$, and the result 
\begin{equation}
\frac{A_{eff}}{A_h} = -\frac{\tilde{I}_Z \tilde{\Phi}_{Iz}(\tilde{a})}{\pi
\tilde{a}^2},
\label{Aeffsup}
\end{equation}
which gives numerically the same values as Eq. (\ref{Aeff}).

\begin{figure}
\includegraphics[width=8cm]{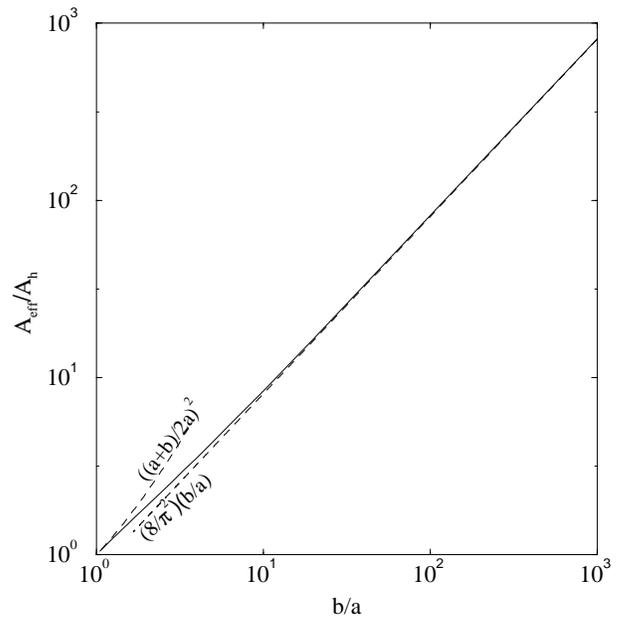}
\caption{%
Reduced effective area, $A_{eff}/A_h$, vs
$1/\tilde{a} = b/a$ for flux focusing calculated from Eq.\
(\ref{Aeff}) or (\ref{Aeffsup}).  Dashed curves show
approximations valid in the limits
$\tilde{a}
\rightarrow 0$ and $\tilde{a} \rightarrow 1.$}
\label{Fig9}
\end{figure}

\section{Geometrical Barrier}
We next present an efficient method for  calculating
the magnetic-field and current-density distributions and the magnetization of a
bulk-pinning-free type-II superconducting disk subject to a geometrical barrier,
which impedes the entry of vortices into the disk.
We consider a disk (radius  $b$ and thickness $d<<b$) in the plane $z=0$,
centered on the $z$ axis, intially in the Meissner state.
 We assume that the London penetration depth obeys $\lambda < d/2$  or, if
$\lambda > d/2$, that the two-dimensional screening length
$\Lambda = 2 \lambda^2/d$ obeys
$\Lambda << b$.
When a perpendicular magnetic field $H_a$ is applied, a sheet-current
density~\cite{Landau60}
\begin{equation}
K_\phi(\rho)=-\frac{4H_a}{\pi}\frac{\rho}{\sqrt{b^2-\rho^2}}
\label{KphiM}
\end{equation}
is induced. 
The resulting magnetic field in the plane $z=0$, determined from Eq.\
(\ref{Hz}), is $H_z(\rho) = 0$ for $\rho<b$ and~\cite{Mikheenko93}
\begin{equation}
H_z(\rho)=H_a\left\{1+\frac{2}{\pi}\left[\frac{1}{\sqrt{(\rho/b)^2-1}}
-\sin^{-1}(\frac{b}{\rho})\right]\right\}
\label{HzM}
\end{equation}
for $\rho>b$.

A geometrical barrier prevents vortices from entering the film until the magnetic
field at the edge (accounting for demagnetizing effects) reaches the value
$H_s$.  We expect that $H_s = H_{c1}$, the lower critical field, if there is no
Bean-Livingston barrier, or $H_s \approx H_c$, the bulk thermodynamic critical
field, if the edge is without defects and thermal activation is negligible. 
An  equivalent criterion is that the magnetic flux begins to penetrate when the
magnitude of the sheet-current density at the edge  reaches the value
$K_s = 2H_s$.~\cite{Kupriyanov75}  To estimate $H_z$ or $K_\phi$  at the edge of
the film, we note that the approximations that led to Eqs. (\ref{KphiM}) and
(\ref{HzM}) break down and that the inverse-square-root divergences in these
equations are cut off when
$\rho$ is within $\delta$ of the edge, where $\delta$ is the larger of $d/2$ or
$\Lambda$. Accordingly, we approximate $H_z$ at the edge of the
film by replacing $\rho$ in the square-root denominator  of Eq.\ (\ref{HzM}) by
$b+\delta$ and using
$\delta << b$, such that $H_z($edge) $\approx (H_a/\pi)\sqrt{2b/\delta}$. 
Similarly, we approximate $K_\phi$ at the edge of the
film by replacing $\rho$ in the square-root denominator  of Eq.\ (\ref{KphiM}) by
$b-\delta$ and using
$\delta << b$, such that $K_\phi($edge) $\approx -(2H_a/\pi)\sqrt{2b/\delta}$. 
Whichever criterion is used [$H_z($edge) $=H_s$ or $|K_\phi($edge$)| = K_s
= 2H_s$], we estimate that the geometrical barrier is overcome when the applied
field is equal to $H_0 = \pi H_s \sqrt{\delta/2b}.$  (In this paper we have
chosen $\tilde{\delta} = \delta / b = 0.01$, such that $H_0 = 0.222 H_s.$  See
Fig. 13.)

When $H_a > H_0$ such that $H_z($edge$) > H_s$, vortices  nucleate at the edge
of the disk and move rapidly towards the center of the disk under the influence
of the Lorentz force per unit length, {\textbf f} = {\textbf J}$_H \times
\vec{\phi}_0$, where {\textbf J}$_H =
\nabla
\times${\textbf H}$_{rev}$,  $\vec{\phi}_0$ is a vector of
magnitude $\phi_0 = h/2e$ along the vortex axis, and {\textbf H}$_{rev}$ is the
thermodynamic magnetic field in equilibrium with the magnetic flux density
{\textbf B} inside the superconductor. As more vortices enter, the return
field outside the disk generated by the vortices inside the disk gradually
reduces the value of the field at the edge to $H_s$, thereby halting
further vortex nucleation.  If bulk pinning is neglible, the case considered
in this paper, the vortices adjust their positions such that the magnetic flux
density (averaged over the  intervortex distance) in the plane
of the disk $B_z(\rho)$ has its maximum value at the center, decreases
monotonically to zero at
$\rho = a$, and remains zero for 
$a < \rho < b$.  The corresponding sheet current density $K_{H\phi}
= J_{H\phi}d $ is zero for
$\rho \le a$, such that the Lorentz force on any vortex vanishes and no further
motion occurs.  Screening supercurrents still flow, however, in the vortex-free
region
$a < \rho < b$.

To good approximation when $d << b$, the resulting magnetic-field and
supercurrent distributions are the same as  those generated by a thin
superconducting annular ring
($a < \rho < b$) in a perpendicular applied field $H_a$, when the solutions are
subject to the constraint that the sheet current density $K_\phi$ is zero at
$\rho = a$.  The Biot-Savart law [Eq.\ (\ref{Hz}) and its extension to $|z| >
0$] guarantees that the current density {\textbf J}$_B =
\nabla
\times${\textbf B}$/\mu_0$ is zero everywhere except within the ring $a < \rho <
b$; thus $K_{B\phi} = J_{B\phi}d$ is zero for $\rho \le a$.  Because {\textbf
J}$_H$ and {\textbf J}$_B$ in thin films are dominated by the curvature of
{\textbf H}$_{rev}$ and  ${\textbf B}/\mu_0$, rather than by the gradients
$\nabla H_{rev}$ and $\nabla B/\mu_0$,\cite{Frankel79,Benkraouda96} it can be
shown that the difference between 
$K_{H\phi}$ and  $K_{B\phi}$ is of order $(d/b)H_a$, decreases for $B > 2
B_{c1}$ as {\textbf H}$_{rev}$ approaches  ${\textbf B}/\mu_0$, and is
negligible for the thin films considered in this paper ($d/b = 0.01$). 
Nevertheless, our simplified approach would be incapable of calculating details
in the
 structure that has been observed in the magnetic flux-density distribution at
the vortex-lattice melting transition.~\cite{Soibel00}  To treat such a problem
would require a more refined approach such as that in Refs.\
\onlinecite{Doyle97} and \onlinecite{Doyle00}, which calculates the local
{\textbf J}$_B$ currents flowing at the vortex solid-liquid interface and
distinguishes between {\textbf H}$_{rev}$ and  ${\textbf B}/\mu_0$.

The magnetic-field and
supercurrent distributions for the case of a thin pin-free disk subject to a
geometrical barrier therefore can be calculated efficiently by using an approach
similar to that used in Secs. II-V.  When a Lorenz-force-free dome of magnetic
flux occupies the region $\rho < a$, the sheet current density in the region $a
< \rho < b$ is $K_{G\phi} = H_a \tilde{K}_{G\phi}$, where the subscript $G$
hencefore labels all quantities that are specific to the geometrical-barrier 
problem.  To evaluate the coefficients $g_{Gm}$ in 
\begin{equation}
  g_G(u) = \sum_{m=1}^{N} g_{Gm}(\frac{u-\tilde{a}}{1-\tilde{a}})^{m-1},
\label{gG}
\end{equation}
we use the $N$ equations
\begin{equation}
  \sum_{m=1}^{N} \alpha_{Fnm}g_{Gm} =  \beta_{Gn},
\label{alphaGsum}
\end{equation}
$n = 1, 2, ..., N$,
where $\alpha_{Gnm} =  \alpha_{nm}$ and $\beta_{Gn} = -1$
for  $n<N$, and 
 $\alpha_{GNm} =  \delta_{1m}$ and $\beta_{GN} = 0$ for
$n=N.$ These equations  are
obtained from  Eqs.\ (\ref{Hz}), (\ref{K}), (\ref{g}), (\ref{hm}), and
(\ref{alphanm}) and
$H_z(\rho_n) = 0$  for
$n<N$, and  from  Eqs.\  (\ref{K}), (\ref{g}), and
$\tilde{K}_{G\phi}(\tilde{a})=0$ for
$n=N$.

\begin{figure}
\includegraphics[width=8cm]{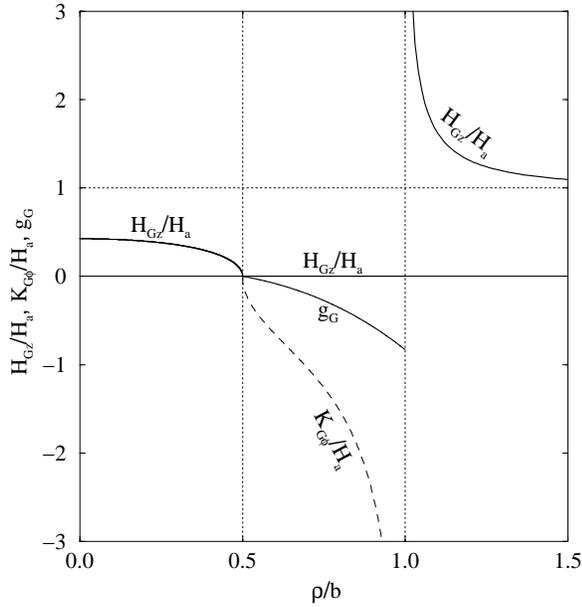}
\caption{%
Reduced magnetic field $\tilde{H}_{Gz} = H_{Gz}/H_a$,
reduced sheet-current density $\tilde{K}_{G\phi}$, and polynomial
$g_G$ vs
$u=\rho/b$ for a pin-free disk of radius $b$ with a geometrical barrier
and a flux dome of reduced radius $\tilde{a} = a/b = 0.5$.}
\label{Fig10}
\end{figure}

Numerical results for $\tilde{H}_{Gz} = H_{Gz}/H_a, \tilde{K}_{G\phi},$ and
$g_G$ vs $u = \rho/b$ for  $a = b/2$ $(\tilde{a} = 0.5)$ are shown in Fig.\ 10. 
In these calculations, we have made no distinction between {\textbf H}$_{rev}$
and 
${\textbf B}/\mu_0$, which corresponds to assuming that $B \approx \mu_0
H$.  However, in cases for which $B$ differs significantly from $\mu_0
H$, our plots of $H_{Gz}/H_a$ (such as in Fig. \ref{Fig10}) would correspond most
closely to plots of the reduced flux density $B_{Gz}/B_a$. 
Results for $g_{Gm}$ vs $\tilde{a}$ are shown in Fig.\ 11.  The magnetic flux
contained within $\rho < a$ can be obtained from
\begin{equation}
\Phi_{Gz}(a) = \mu_o H_a b^2
\tilde{\Phi}_{Gz}(\tilde{a}),
\label{PhiGz}
\end{equation}
where
\begin{equation}
\tilde{\Phi}_{Gz}(\tilde{a}) =\pi \tilde{a}^2 + 
\sum_{m=1}^{N} \phi_m(\tilde{a}) g_{Gm},
\label{PhitildeGz}
\end{equation}
and the average magnetic flux density in the disk is $B_{av} =\Phi_{Gz}(a)/\pi
b^2$. Figure 12 shows how  $B_{av}/B_a$ and $H_{Gz}(0)/H_a$,
where
$H_{Gz}(0)$ is the magnetic field at the center of the disk, depend upon  
$\tilde{a}$.

\begin{figure}
\includegraphics[width=8cm]{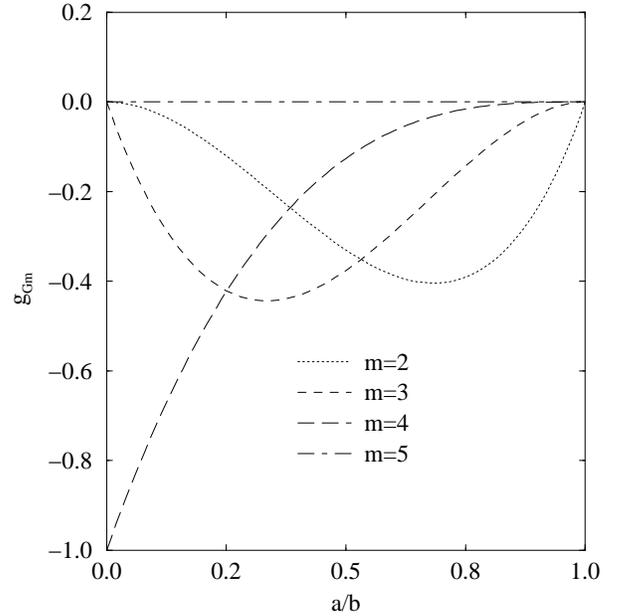}
\caption{%
Coefficients $g_{Gm}$ in the polynomial of Eq.\ (\ref{gG}) vs $\tilde{a}$ for a
pin-free disk of radius $b$ with a geometrical barrier and a flux dome of
reduced radius $\tilde{a} = a/b$.    We require $g_{G1}=0$.}
\label{Fig11}
\end{figure}

\begin{figure}
\includegraphics[width=8cm]{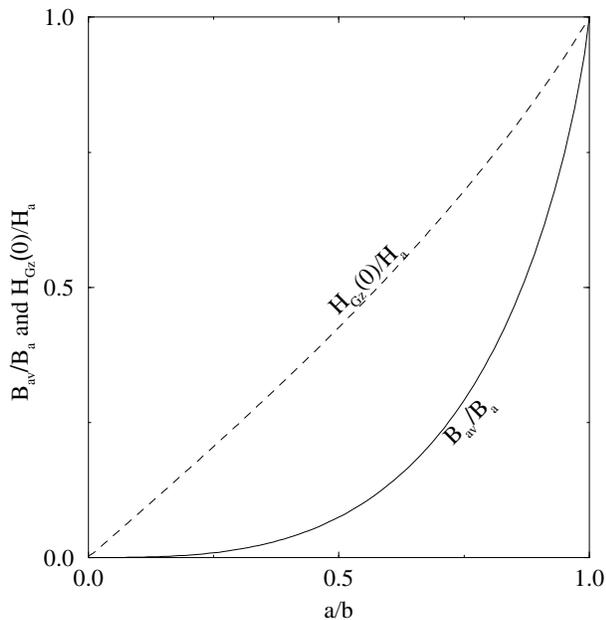}
\caption{%
Reduced average flux density $B_{av}/B_a$ (solid) and reduced flux density at
the center $H_{Gz}(0)/H_a$ (dashed) of a pin-free disk of radius $b$ with a
geometrical barrier and a flux dome of reduced radius $\tilde{a} = a/b$.}
\label{Fig12}
\end{figure}

We next calculate  the magnetization, i.e., the magnetic moment per unit volume
of the disk,
$M_{Gz} = m_{Gz}/\pi b^2d$, where $m_{Gz}$ is calculated from Eq.\ (\ref{mz}).
The initial magnetization of the disk in the Meissner state ($0 \le H_a \le
H_0$, see Fig.\ 13), calculated from Eq.\ (\ref{KphiM}), is~\cite{Clem94} $M_{Gz}
= -\chi_0 H_a$, where $\chi_0 = 8b/3\pi d$; i.e., the external magnetic
susceptibility~\cite{Goldfarb92} in this case is $\chi = -\chi_0$.  Whenever
there is a dome of magnetic flux within the region $\rho < a$,  the
magnetization, obtained from Eqs.\ (\ref{mz}), (\ref{K}),
(\ref{fm}), and (\ref{gG}), may be calculated from
\begin{equation}
M_{Gz} = \frac{3\pi}{8}\chi_0 H_a 
\sum_{m=1}^{N} f_m g_{Gm},
\label{MGz}
\end{equation}
where $f_m$ and $g_{Gm}$ depend
implicitly upon $\tilde{a}$.  

\begin{figure}
\includegraphics[width=8cm]{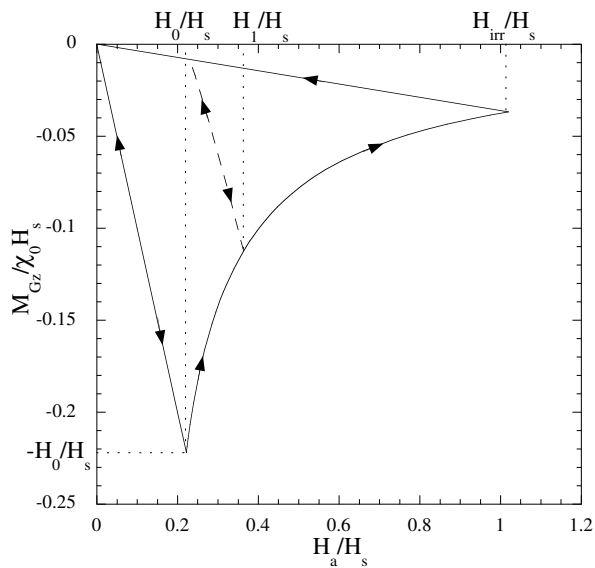}
\caption{%
Calculated hysteresis in the reduced magnetization $M_{Gz}/\chi_0 H_s$ vs reduced
applied field
$H_a/H_s$ for a pin-free disk with a geometrical barrier (solid).  The dashed
curve shows a reversible minor hysteresis ``loop" occurring when the applied
field is reduced after the applied field $H_a$ has reached 
$H_1$ along the field-increasing magnetization curve.  As $H_a$ decreases, the
flux dome expands, but the flux contained within the dome remains constant.}
\label{Fig13}
\end{figure}

For $H_0 <
H_a < H_{irr}$ along the field-increasing magnetization curve at the critical
entry condition (see Fig.\ 13), $H_a$ and $\tilde{a}$ are related via 
\begin{equation}
H_a = -H_0 \sqrt{1-\tilde{a}^2} /\sum_{m=1}^{N} g_{Gm},
\label{Haentry}
\end{equation}
This equation follows from the condition 
that $|K_{G\phi}($edge$)|=2H_s$,
where $K_{G\phi}($edge) is obtained by evaluating Eq.\ (\ref{K}) at $u=1$ but
replacing $\sqrt{1-u^2}$ in the denominator by $\sqrt{2\delta/b}$, as in the
evaluation of
$H_0$. When $\tilde{a}= 0$, we see by comparing Eqs.\ (\ref{K}) and
(\ref{KphiM}) that  $g_G(u) = -u^3,$ such that $g_{Gm} = -\delta_{m4}$ (see also
Fig.\ 11),
$f_4 = 8/3\pi$, and $M_{Gz\uparrow} = -\chi_0 H_0$ at $H_a = H_0$.  In the limit
as
$\tilde{a} \rightarrow 1$, $\tilde{K}_{G\phi} \approx -2 \sqrt{u-a}/\sqrt{1-u}$,
such that $g_G(u) \approx -\pi(u-\tilde{a})$, $g_{Gm} \approx -\pi
(1-\tilde{a}) \delta_{m2}$, $f_2 \approx 1$, $H_a \approx H_0 \sqrt{2}/\pi
\sqrt{1-\tilde{a}}$, and $M_{Gz\uparrow}/\chi_0 H_s \approx -(3
\pi^2/8) \tilde{\delta} H_s/H_a$,
where $\tilde{\delta} = \delta / b << 1$. 

For $H_0 < H_a < H_{irr}$ along the field-decreasing magnetization curve at the
critical exit condition, we assume that the  radius $a$  of the vortex-filled
region has reached within
$\delta$ of the radius $b$ of the disk; i.e., $\tilde{a}=1-\tilde{\delta}$.  Using Eq.\ (\ref{MGz}) with  
$g_{Gm} \approx -\pi
\tilde{\delta} \delta_{m2}$ and $f_2 \approx 1$, we obtain 
$M_{Gz\downarrow}/\chi_0 H_s \approx -(3
\pi^2/8) \tilde{\delta} H_a/H_s$.  See Fig.\ 13.

The field-increasing and field-decreasing magnetization curves in Fig.\ 13 meet
at $H_a = H_{irr}$, the irreversibility field.  The criteria we used for the
critical entry and exit conditions lead to the result that $H_{irr} \approx
H_s$, where the magnetization is given by $M_{Gzirr}/\chi_0 H_s 
\approx -(3 \pi^2/8) \tilde{\delta}$.  However, the above expressions for
$H_{irr}$, 
$M_{Gzirr}$, and  $M_{Gz\downarrow}$ are the least reliable results of our
paper, because all these quantities are very sensitive to the precise conditions
for entry and exit at the edge of the disk, including such details as the shape
of the edge.~\cite{Schuster94,Zeldov94a,Benkraouda96,Mawatari03,
Doyle97,Doyle00}  The magnetic moment responsible for the magnetization
$M_{Gzirr}$ and $M_{Gz\downarrow}$ is produced by currents that flow only within
a very narrow band around the disk's edge, where a theory more accurate than
ours is needed.  

The minor hysteresis ``loop," shown as the dashed curve in Fig.\ 13, can be
calculated as follows.  We start at a point on the field-increasing magnetization
curve where the flux dome has radius $a_1$.  The magnetic flux
contained within the dome $\Phi_{Gz}(a_1)$, the magnetization
$M_{Gz1}$, and the corresponding applied field  $H_1$ are obtained from Eqs.\
(\ref{PhiGz}), (\ref{MGz}), and (\ref{Haentry}), where 
$\tilde{a}$, $f_m$, and
$g_{Gm}$ are all evaluated at $\tilde{a} = \tilde{a}_1 = a_1/b$.  As the applied
field $H_a$ is reduced from its starting value $H_1$, the radius $a$ of the flux
dome expands, but the magnetic flux within
the dome  remains constant.  For each value of $\tilde{a} >
\tilde{a}_1$, we recalculate $f_m$, 
$g_{Gm}$, and $\tilde{\Phi}_{Gz}(\tilde{a})$.  We then  use Eq.\ (\ref{PhiGz}) to
obtain the corresponding value of the applied field,
\begin{equation}
H_a = H_1 \tilde{\Phi}_{Gz}(\tilde{a}_1)/ \tilde{\Phi}_{Gz}(\tilde{a}),
\label{Haminor}
\end{equation}
and Eq.\ (\ref{MGz}) to obtain the corresponding value of the magnetization.

\section{Discussion}
In this paper, we have presented an efficient method for the calculation of
magnetic-field and current-density profiles for thin-film rings in the
Meissner state and for bulk-pinning-free disks subject to a geometrical
barrier.  In each case, the sheet-current density was expressed in the form of
Eq.\ (\ref{K}), where the quantity $g(u)$ in the numerator is a polynomical of
degree $N-1$.  

For all the calculations presented in the figures, for which we
assumed
$N = 5$, we found that the magnitude of $g_5$ was less than 0.0012 in each
case (see Figs.\ 2, 5, 8, and 11) and that its contribution to $g(u)$ was less
than 1.1\%.  Using $N = 6$ yields values of $g_6$ whose magnitudes are much
smaller than those of $g_5$, and the values of the calculated physical
quantities are altered only in the sixth decimal place.  

Moreover, we offer the conjecture
that the problems we solved numerically in Secs. III-VI might be solved
analytically with functions $g_I, g_Z, g_F,$ and $g_G$ that are third-order
polynomials in $u$; i.e., the sums in Eqs.\ (\ref{g}), (\ref{gI}), (\ref{gZ}),
(\ref{gF}), and  (\ref{gG}) might simply terminate with $N=4$.  As evidence in
support of this conjecture, we note that our calculations for $\tilde{a} = a/b =
0.1$ and 0.5 with $N$ = 4, 5, 6, and 7 yielded values of $L/\mu_0 b$
[Eqs.\ (\ref{L}) and (\ref{PhiIz})] that differed
only in the fifth decimal place.
Similarly, values of $I_Z/H_ab$ [Eq.\ (\ref{IZ})], $A_{eff}/A_h$ [Eq.\
(\ref{Aeff})], and $M_{Gz\uparrow}/\chi_0H_s$ [Eqs.\ (\ref{MGz}) and
(\ref{Haentry})] calculated for $\tilde{a} = a/b =
0.1$ and 0.5 with $N$ = 4, 5, 6, and 7 differed at most only in the fourth
significant figure. It is possible that the values we obtained for $g_5, g_6,$
and
$g_7$ in Secs. III-VI were nonzero only because of small numerical errors 
introduced because we performed the integrals in Eqs.\ (\ref{hm})-(\ref{phim})
numerically rather than analytically.

Although in this paper we have considered only bulk-pinning-free  thin-film rings
and disks, it should be possible to extend the present approach
to develop an efficient method, complementary to that of Ref.\
\onlinecite{Gilchrist96}, for numerically calculating quasistatic magnetic-field
and current-density distributions in rings and disks subject to both a
geometrical barrier and bulk pinning.  Such distributions recently have been
calculated analytically for infinitely long strips in Refs.\
\onlinecite{Maksimov98,Elistratov00,Zhelezina02,Maksimov02,Elistratov02}.
 
\begin{acknowledgments}
We thank J. Clarke and V. G. Kogan for stimulating discussions.  This manuscript
has been authored in part by Iowa State University of Science  and Technology
under Contract No.\ W-7405-ENG-82 with the U.S.\ Department of  Energy.
\end{acknowledgments}



\begin{thebibliography}{99}
\bibitem{footnote} Similarly, the inductance per unit length [Eq. (17) of
Ref.~\onlinecite{Babaei02}]
of a pair of parallel coplanar superconducting strips of 
center-to-center distance $s$ and
strip-width $w$ approaches
$L'_0 =  (\mu_0/\pi)\ln(4s/w)$ when $w << s$, which
can be obtained from $L' =
(\mu_0/\pi)\ln(s/r)$, the inductance per unit length of a pair of
superconducting wires of center-to-center distance $s$ and radius $r$, by
replacing $r$ by
$w/4$.
\end{thebibliography}
\end{document}